\documentclass[fleqn,twoside]{article}
\usepackage[headings]{espcrc2}

\readRCS
$Id: espcrc2.tex,v 1.2 2004/02/24 11:22:11 spepping Exp $
\usepackage{graphicx}
\usepackage[figuresright]{rotating}

\newcommand{\AmS}{{\protect\the\textfont2
  A\kern-.1667em\lower.5ex\hbox{M}\kern-.125emS}}

\newcommand{\lsim}
{\mathrel{\raisebox{-.3em}{$\stackrel{\displaystyle <}{\sim}$}}}

\def\asymp#1%
{\mathrel{\raisebox{-.4em}{$\widetilde{\scriptstyle #1}$}}}

\def\Nequal#1%
{\mathrel{\raisebox{-.5em}{$\stackrel{=}{\scriptstyle\rm#1}$}}}
\newcommand{\dsl}[1]{\not \hspace{-0.7mm}#1}
\def\dsl{\mathpalette\make@slash}
\def\make@slash#1#2{\setbox\z@\hbox{$#1#2$}%
  \hbox to 0pt{\hss$#1/$\hss\kern-\wd0}\box0}

\def\beq{\begin{equation}}
\def\eeq{\end{equation}}
\def\beqar{\begin{eqnarray}}
\def\eeqar{\end{eqnarray}}
\def\barr#1{\begin{array}{#1}}
\def\earr{\end{array}}
\def\bfi{\begin{figure}}
\def\efi{\end{figure}}
\def\btab{\begin{table}}
\def\etab{\end{table}}
\def\bce{\begin{center}}
\def\ece{\end{center}}

\def\text{\textstyle}

\def\de{\delta}

\def\si{\sigma}

\def\reffi#1{\mbox{Figure~\ref{#1}}}

\def\citere#1{\mbox{Ref.~\cite{#1}}}
\def\citeres#1{\mbox{Refs.~\cite{#1}}}

\newcommand{\TeV}{\unskip\,\mathrm{TeV}}
\newcommand{\GeV}{\unskip\,\mathrm{GeV}}
\newcommand{\MeV}{\unskip\,\mathrm{MeV}}

\newcommand{\Oa}{\mathswitch{{\cal{O}}(\alpha)}}

\def\mathswitchr#1{\relax\ifmmode{\mathrm{#1}}\else$\mathrm{#1}$\fi}

\newcommand{\PW}{\mathswitchr W}

\newcommand{\Pe}{\mathswitchr e}

\newcommand{\Pd}{\mathswitchr d}
\newcommand{\Pdbar}{\bar{\mathswitchr d}}
\newcommand{\Pu}{\mathswitchr u}

\newcommand{\Ps}{\mathswitchr s}

\newcommand{\Pc}{\mathswitchr c}

\newcommand{\Pep}{\mathswitchr {e^+}}
\newcommand{\Pem}{\mathswitchr {e^-}}
\newcommand{\PWp}{\mathswitchr {W^+}}

\def\mathswitch#1{\relax\ifmmode#1\else$#1$\fi}

\newcommand{\MW}{\mathswitch {M_\PW}}

\newcommand{\GW}{\Gamma_{\PW}}

\def\solid{\raise.9mm\hbox{\protect\rule{1.1cm}{.2mm}}}
\def\dash{\raise.9mm\hbox{\protect\rule{2mm}{.2mm}}\hspace*{1mm}}

\def\ie{i.e.\ }

\newcommand{\eefourf}{{\mathswitchr{ee4f}}}

\title{Electroweak Corrections to $\Pep\Pem\to$ 4 fermions}

\author{A.~Denner\address[PSI]{Paul Scherrer Institut, 
                               W\"urenlingen und Villigen,
        CH-5232 Villigen PSI, Switzerland}, 
        S.~Dittmaier\address[MPI]{Max-Planck-Institut f\"ur Physik
        (Werner-Heisenberg-Institut),
        D-80805 M\"unchen, Germany},
M. Roth\addressmark[MPI], L.H. Wieders\addressmark[PSI]}
       
\runtitle{Electroweak Corrections to Four-Fermion Production in
$\Pep\Pem$ Annihilation}
\runauthor{A. Denner, S. Dittmaier, M. Roth, L.H. Wieders}

\begin{document}

\begin{abstract}
  The calculation of the full electroweak ${\cal O}(\alpha)$
  corrections to the charged-current four-fermion production processes
  $\Pep\Pem\to\nu_\tau\tau^+\mu^-\bar\nu_\mu$,
  $\Pu\bar\Pd\mu^-\bar\nu_\mu$, and $\Pu\bar\Pd\Ps\bar\Pc$ is briefly
  reviewed.  The calculation is performed using the complex-mass
scheme for the gauge-boson resonances.  The evaluation of the
occurring one-loop tensor integrals, which include 5- and 6-point
functions, requires new techniques.  The effects of the complete $\Oa$
corrections to the total cross section and to the production-angle
distribution are discussed and compared to predictions based on the
double-pole approximation, revealing that the latter approximation is
not sufficient to fully exploit the potential of a future linear
collider in an analysis of W-boson pairs at high energies.
\vspace{1pc}
\end{abstract}

\maketitle

\section{INTRODUCTION}

At LEP2, W-pair-mediated four-fermion $(4f)$ production was
experimentally explored with quite high precision (see \citere{lep2}
and references therein).  For LEP2 accuracy, it was sufficient to
include corrections in the so-called double-pole approximation (DPA),
where only the leading term in an expansion about the poles in the two
W-boson propagators is taken into account.  Different versions of such
a DPA have been used in the literature
\cite{Beenakker:1998gr,Jadach:1998tz,Jadach:2000kw,Denner:2000kn,Kurihara:2001um}.
Although several Monte Carlo programs exist that include universal
corrections, only two event generators, {\sc YFSWW}
\cite{Jadach:1998tz,Jadach:2000kw} and {\sc RacoonWW}
\cite{Denner:2000kn,Denner:1999gp,Denner:2001zp}, include
non-universal corrections.

In the DPA approach, the W-pair cross section can be predicted within
$\sim0.5\%$ $(0.7\%)$ in the energy range between $180\GeV$
($170\GeV$) and $\sim 500\GeV$, which was sufficient for the LEP2
accuracy of $\sim 1\%$. In the threshold region ($\sqrt{s}\lsim
170\GeV$), the DPA is not reliable, and the best available prediction
resulted from an improved Born approximation (IBA) based on leading
universal corrections only, and thus possesses an intrinsic
uncertainty of $\sim 2\%$.

At a future International $\Pe^+ \Pe^-$ Linear Collider (ILC)
\cite{Aguilar-Saavedra:2001rg,Abe:2001wn,Abe:2001gc}, the accuracy of
the cross-section measurement will be at the per-mille level, and the
precision of the $\PW$-mass determination is expected to be
\mbox{$\sim 7\MeV$} from a threshold scan of the total
W-pair-production cross section
\cite{Aguilar-Saavedra:2001rg,Abe:2001wn}.  For the cross-section
prediction at threshold the theoretical uncertainty of the IBA is
$\sim 2\%$ and thus definitely insufficient for the planned precision
measurement of $\MW$ in a threshold scan.

Recently we have completed the first calculation of the complete
${\cal O}(\alpha)$ corrections (improved by higher-order ISR) for
$\Pep\Pem\to\nu_\tau\tau^+\mu^-\bar\nu_\mu$,
$\Pu\bar\Pd\mu^-\bar\nu_\mu$, and $\Pu\bar\Pd\Ps\bar\Pc$, which are
relevant for W-pair production.  We have presented results on total
cross sections in \citere{Denner:2005es} and on various differential
distributions in \citere{Denner:2005fg}. The latter publication also
contains technical details of the calculation, which is rather
complicated.%
\footnote{Some problems related to the finite width that can appear in
  such a calculation are illustrated in \citere{Boudjema:2004id}.}  In
the following we briefly describe the salient features of the
calculation and present a selection of numerical results that are
relevant for a future ILC.

\section{METHOD OF CALCULATION}

The actual calculation builds upon the {\sc RacoonWW} approach
\cite{Denner:2000kn}, where real-photonic corrections are based on
full matrix elements and virtual corrections are treated in DPA. 
Real and virtual corrections are combined either using two-cutoff
phase-space slicing or employing the dipole subtraction method
\cite{Dittmaier:2000mb} for photon radiation.  

In contrast to the DPA approach, the one-loop calculation of an
$\Pep\Pem\to4f$ process requires the evaluation of 5- and 6-point
one-loop tensor integrals.  We calculate the 6-point integrals by
directly reducing them to six 5-point functions, as described in
\citeres{Denner:1993kt,Denner:2005nn}.  The 5-point integrals are
reduced to five 4-point functions following the methods of
\citeres{Denner:2005nn,Denner:2002ii}. Note that this reduction of 5-
and 6-point integrals to 4-point integrals does not involve inverse
Gram determinants composed of external momenta, which naturally occur
in the Passarino--Veltman reduction \cite{Passarino:1979jh} of tensor
to scalar integrals. The latter procedure leads to serious numerical
problems when the Gram determinants become small.

Tensor 4-point and 3-point integrals are reduced to scalar integrals
with the Passarino--Veltman algorithm \cite{Passarino:1979jh} as long
as no small Gram determinant appears in the reduction. If small Gram
determinants occur, two alternative schemes are applied
\cite{Denner:2005nn}.  One method makes use of expansions of the
tensor coefficients about the limit of vanishing Gram determinants and
possibly other kinematical determinants. In this way, again all tensor
coefficients can be expressed in terms of the standard scalar
functions. In the second, alternative method we evaluate a specific
tensor coefficient, the integrand of which is logarithmic in Feynman
parametrization, by numerical integration. Then the remaining
coefficients as well as the standard scalar integral are algebraically
derived from this coefficient.

As a further complication, also the evaluation of the three spinor
chains corresponding to the three external fermion--antifermion pairs
is non-trivial, because the chains are contracted with each other
and/or with four-momenta in many different ways. There are ${\cal
  O}(10^3)$ different chains to calculate.  We have worked out
algorithms that algebraically reduce all these spinor chains to a few
or, in an alternative method, to a minimal set of standard structures
without introducing numerical problems.  These 
\looseness -1
algorithms are described in \citere{Denner:2005fg} in detail.

The description of resonances in (standard) perturbation theory
requires a Dyson summation of self-energy insertions in the resonant
propagator in order to introduce the imaginary part provided by the
finite decay width into the propagator denominator.  This procedure in
general violates gauge invariance, \ie destroys Slavnov--Taylor or
Ward identities and disturbs the cancellation of gauge-parameter
dependences, because different perturbative orders are mixed (see, for
instance, \citere{Grunewald:2000ju} and references therein).

For our calculation we have generalized \cite{Denner:2005fg} the
so-called ``complex-mass scheme'', which was introduced in
\citere{Denner:1999gp} for lowest-order calculations, to the one-loop
level.  In this approach the W- and Z-boson masses are consistently
considered as complex quantities, defined as the locations of the
propagator poles in the complex plane.  To this end, bare real masses
are split into complex renormalized masses and complex counterterms.
Since the bare Lagrangian is not changed, double counting does not
occur.  Perturbative calculations can be performed as usual, only
parameters and counterterms, in particular the electroweak mixing
angle defined from the ratio of the W- and Z-boson masses, become
complex. Since we only perform an analytic continuation of the
parameters, all relations that follow from gauge invariance, such as
Ward identities, remain valid. As a consequence the amplitudes are
gauge independent, and unitarity cancellations are respected.
Moreover, the on-shell renormalization scheme can straightforwardly be
transferred to the complex-mass scheme \cite{Denner:2005fg}.

The use of complex gauge-boson masses necessitates the consistent use
of these complex masses also in loop integrals.  The scalar master
integrals are evaluated for complex masses using the methods and
results of Refs.~\cite{'tHooft:1978xw,Beenakker:1988jr,Denner:1991qq}.

In order to prove the reliability of our results, we have carried out
various checks, as described in detail in \citere{Denner:2005es}.
We have checked the structure of the (UV, soft, and collinear)
singularities, the matching between virtual and real corrections, and
the gauge independence (by performing the calculation in the
't~Hooft--Feynman gauge and in the background-field gauge
\cite{Denner:1994xt}).  
\looseness -1 
The most convincing check for ourselves is the fact that we worked out
the whole calculation in two independent ways, resulting in two
independent computer codes the results of which are in good agreement.

\section{NUMERICAL RESULTS}

The precisely defined input for the numerical results presented in the
following can be found in \citeres{Denner:2005es,Denner:2005fg}.

\begin{figure*}%
{\unitlength 1cm
\begin{picture}(16,14.2)
\put(-4.6,- 8.2){\includegraphics{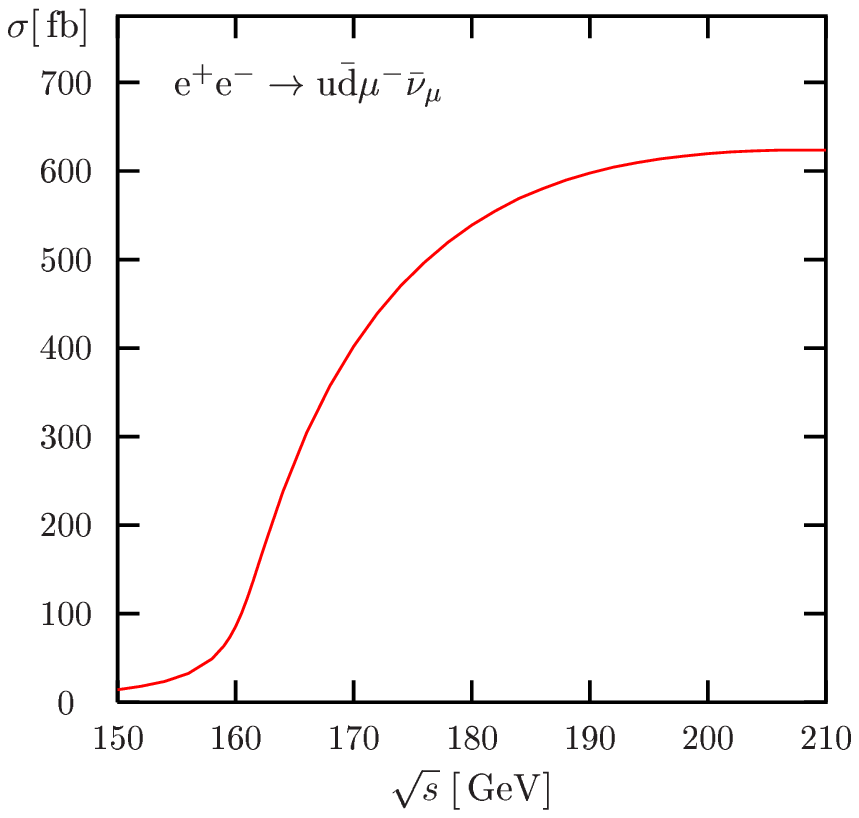}}
\put( 3.6,- 8.2){\includegraphics{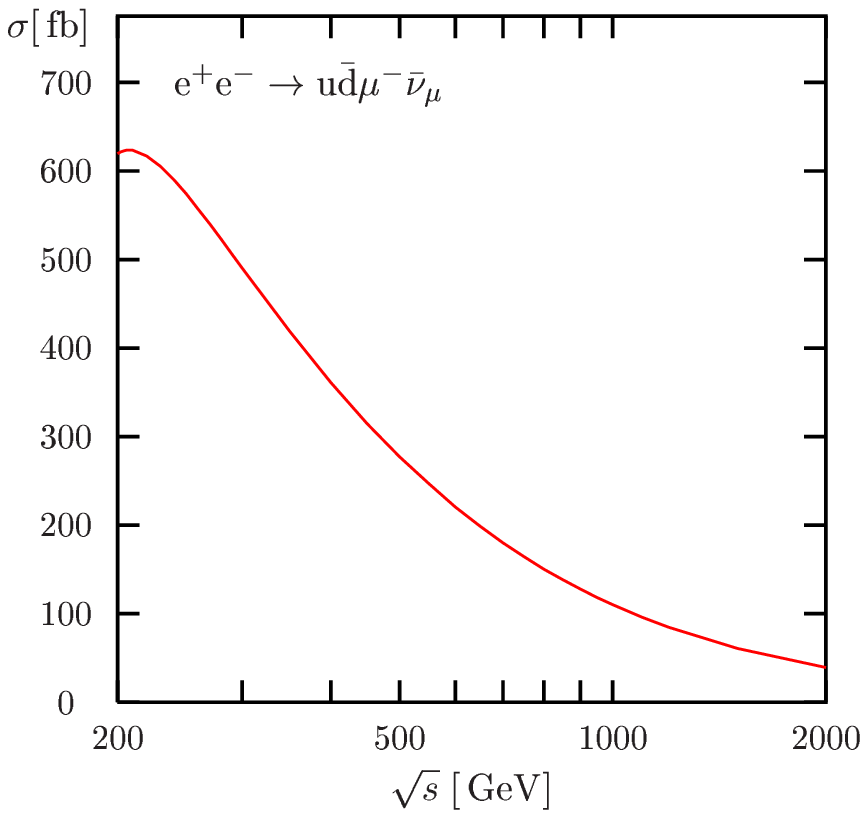}}
\put(-4.6,-16.2){\includegraphics{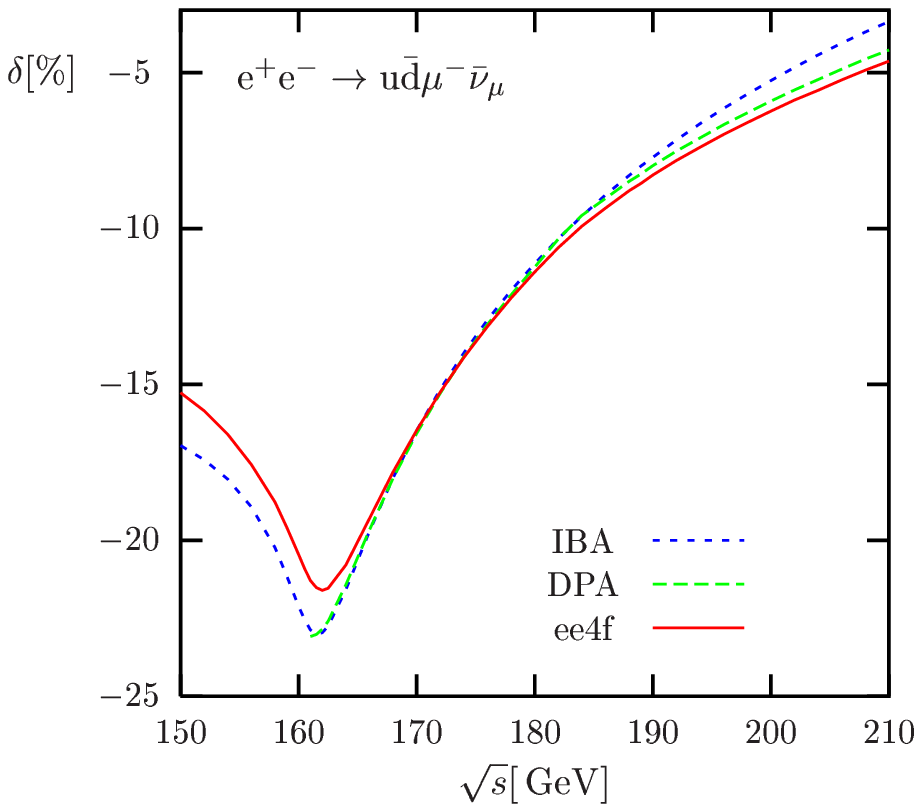}}
\put( 3.6,-16.2){\includegraphics{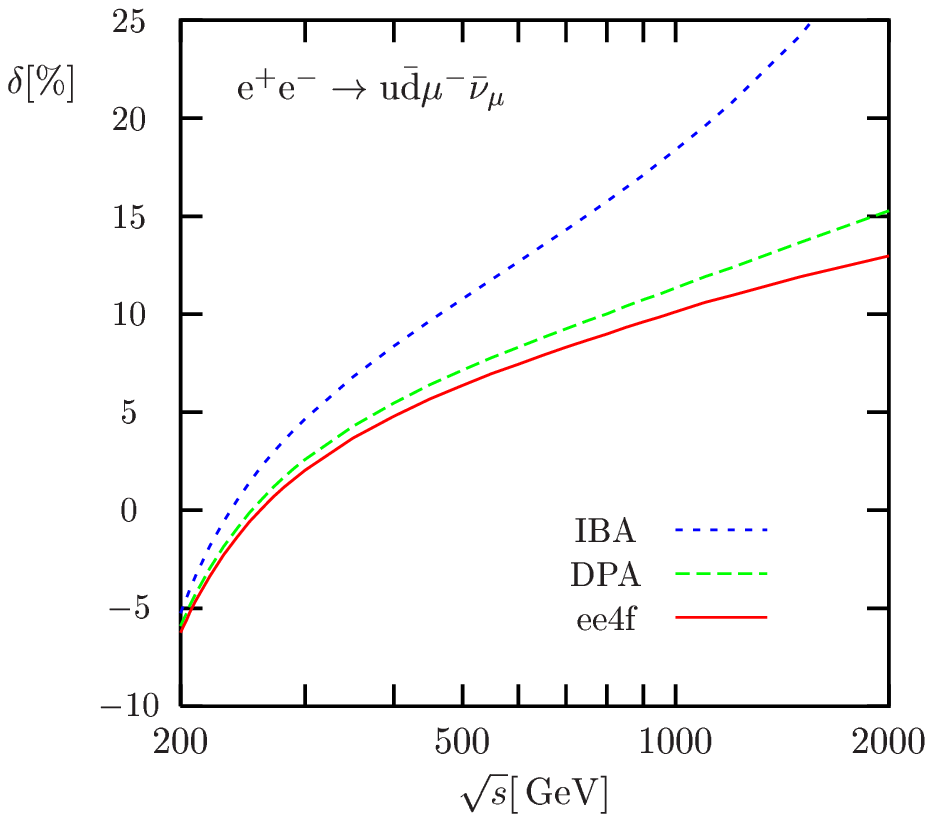}}
\end{picture} }
\caption{Absolute cross section $\si$
(upper plots) and relative corrections $\de$ (lower plots)
to the total cross section without cuts for
$\Pep\Pem\to\Pu\Pdbar\mu^-\bar\nu_\mu$ obtained from the
IBA, DPA, and the full ${\cal O}(\alpha)$ calculation (\eefourf). All
predictions are improved by higher-order ISR. 
}
\label{fig:leptrelrc}
\end{figure*}

Figure~\ref{fig:leptrelrc} depicts the total cross section
for the energy ranges of LEP2 and of the high-energy phase of a future ILC,
focusing on the semileptonic final state $\Pu\Pdbar\mu^-\bar\nu_\mu$.
The respective figures for the relative corrections $\de$ to the
leptonic (shown in \citere{Denner:2005es}) and hadronic final states
look almost identical, up to an offset resulting from the missing or
additional QCD corrections.  Specifically, the upper plots show the
absolute prediction for the cross section including the full $\Oa$
corrections and improvements from higher-order ISR.  The lower plots
compare the relative corrections as obtained from the full $\Oa$
calculation, from an IBA, and from the DPA.  The IBA
\cite{Denner:2001zp} implemented in {\sc RacoonWW} is based on
universal corrections only and includes solely the contributions of
the CC03 diagrams.  The DPA of {\sc RacoonWW} comprises also
non-universal corrections \cite{Denner:2000kn} and goes beyond a pure
pole approximation in two respects. The real-photonic corrections are
based on the full $\Pep\Pem\to4f+\gamma$ matrix elements, and the
Coulomb singularity is included for off-shell W~bosons.  Further
details can be found in \citere{Denner:2000kn}.

A comparison between the DPA and the predictions based on the full
${\cal O}(\alpha)$ corrections reveals differences in the relative
corrections $\de$ of $\lsim 0.5\%\; (0.7\%)$ for CM energies ranging
from $\sqrt{s}\sim170\GeV$ to $300\GeV\; (500\GeV)$.  This is in
agreement with the expected reliability of the DPA, as discussed in
\citeres{Jadach:2000kw,Denner:2000kn,Grunewald:2000ju}.  At higher
energies, the deviations increase and reach $1{-}2\%$ at
$\sqrt{s}=1{-}2\TeV$.  In the threshold region
($\sqrt{s}\lsim170\GeV$), as expected, the DPA also becomes worse
w.r.t.\ the full one-loop calculation, because the naive error
estimate of $({\alpha}/{\pi})({\GW}/{\MW})$ times some numerical
safety factor of ${\cal O}(1{-}10)$ for the corrections missing in the
DPA has to be replaced by $({\alpha}/{\pi}){\GW}/{(\sqrt{s}-2\MW})$
and thus becomes large.  In view of that, the DPA is even surprisingly
good near threshold.  For CM energies below $170\GeV$ the LEP2 cross
section analysis was based on approximations like the shown IBA, which
follows the full one-loop corrections even below the threshold at
$\sqrt{s}=2\MW$ within an accuracy of about $2\%$, as expected in
\citere{Denner:2001zp}.  More results on total cross sections,
including numbers on leptonic, semileptonic, and hadronic final
states, can be found in \citere{Denner:2005es}.

\begin{figure*}
{\unitlength 1cm
\begin{picture}(16,7.5)
\put( -4.9,-18.2){\includegraphics{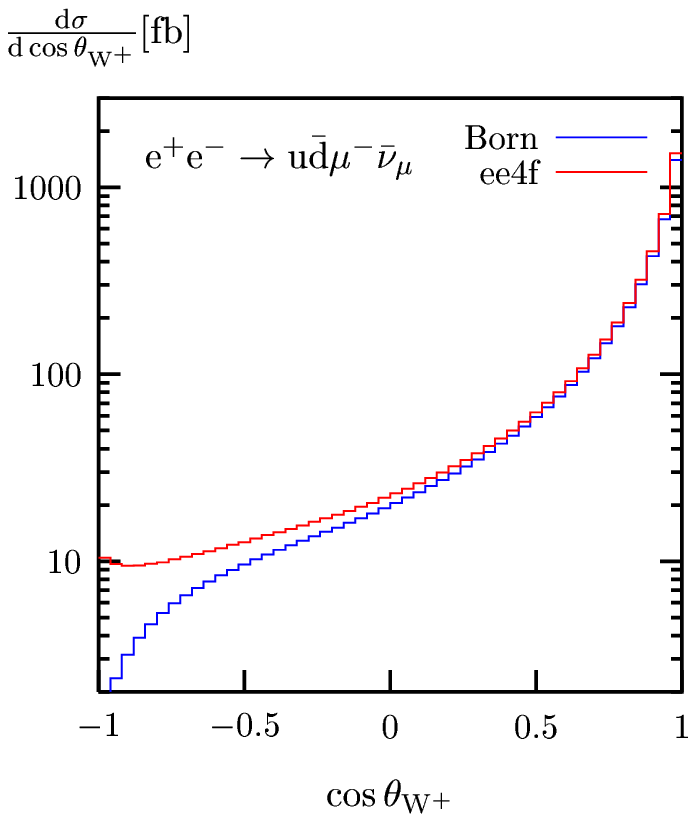}}
\put( 3.3,-18.2){\includegraphics{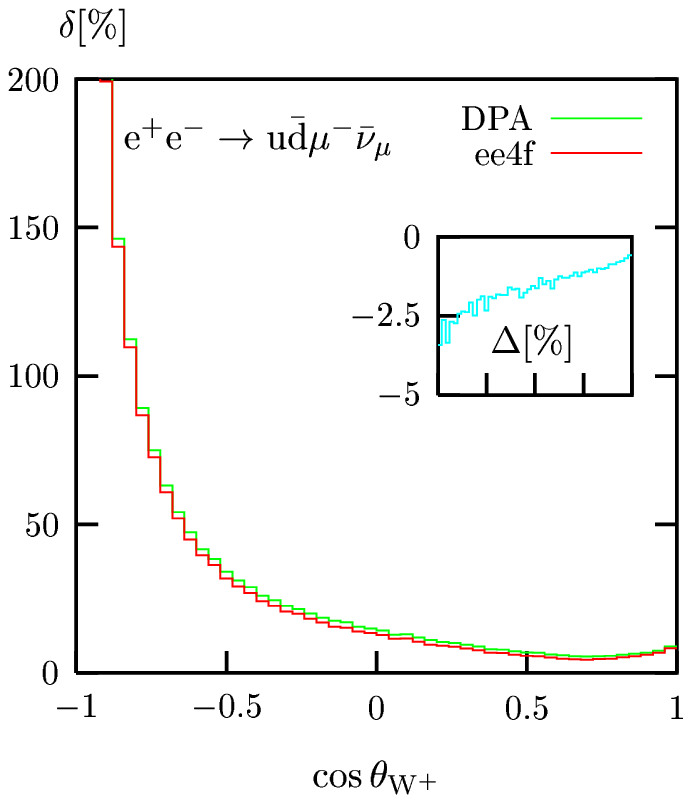}}
\put(3,7.0){$\sqrt{s}=500\GeV$}
\put(11.2,7.0){$\sqrt{s}=500\GeV$}
\end{picture}}%
\caption{Distribution in  the cosine of the  $\protect\PWp$ production angle
  with respect to the $\protect\Pep$ beam (r.h.s.) and the
  corresponding corrections (lower row) at $\sqrt{s}=500\GeV$ 
  for $\Pep\Pem\to \Pu \Pdbar \mu^-
  \bar\nu_{\mu}$. The inset plot shows the relative difference $\Delta$
  between the full
  $\Oa$ corrections and those in DPA. (Taken from \citere{Denner:2005fg}.)}
\label{fi:dist}
\end{figure*}

The distribution in the cosine of the $\PWp$ production angle is shown
in \reffi{fi:dist} for the process $\Pep\Pem\to \Pu \Pdbar \mu^-
\bar\nu_{\mu}$ at $\sqrt{s}=500\GeV$.
Further distributions, also for $\sqrt{s}=200\GeV$, are presented in
\citere{Denner:2005fg}.  For the W-production-angle distribution the
full $\Oa$ calculation and the DPA agree within $\sim 1\%$ for LEP2
energies (see Fig.~12 of \citere{Denner:2005fg}), but at $500\GeV$ the
difference of the corrections in DPA and the complete $\Oa$
corrections rises from $-1\%$ to about $-2.5\%$ with increasing
scattering angle (see inset in r.h.s.\ of \reffi{fi:dist}). Note that
such a distortion of the shape of the angular distribution can be a
\looseness -1 signal for anomalous triple gauge-boson couplings.

\vspace*{2ex}

This work was supported in part by the Swiss National Science
Foundation.


\end{document}